\documentclass[conference, a4paper]{IEEEtran}

\usepackage{cite}
\usepackage{amsmath,amssymb,amsfonts}

\DeclareMathOperator*{\argmin}{arg\,min}
\usepackage{algorithm}
\usepackage{algpseudocode}
\usepackage{graphicx}
\usepackage{textcomp}
\usepackage{xcolor}
\usepackage[caption=false,font=footnotesize]{subfig}
\usepackage{epstopdf}
\def\BibTeX{{\rm B\kern-.05em{\sc i\kern-.025em b}\kern-.08em
    T\kern-.1667em\lower.7ex\hbox{E}\kern-.125emX}}
\begin{document}

\title{Realistic Channel Models Pre-training\\
}

\author{
    \IEEEauthorblockN{Yourui Huangfu\IEEEauthorrefmark{1}, Jian Wang\IEEEauthorrefmark{1}, Chen Xu\IEEEauthorrefmark{1}, Rong Li\IEEEauthorrefmark{1}, Yiqun Ge\IEEEauthorrefmark{2}, Xianbin Wang\IEEEauthorrefmark{1}, Huazi Zhang\IEEEauthorrefmark{1}, Jun Wang\IEEEauthorrefmark{1}}
    \IEEEauthorblockA{\IEEEauthorrefmark{1}Hangzhou Research Center, Huawei Technologies, Hangzhou, China}
    \IEEEauthorblockA{\IEEEauthorrefmark{2}Ottawa Research Center, Huawei Technologies, Ottawa, Canada}
    Emails: \{huangfuyourui,wangjian23,xuchen14,lirongone.li,yiqun.ge,wangxianbin1,zhanghuazi,justin.wangjun\}@huawei.com
}

\maketitle

\begin{abstract}
In this paper, we propose a neural-network-based realistic channel model with both the similar accuracy as deterministic channel models and uniformity as stochastic channel models. To facilitate this realistic channel modeling, a multi-domain channel embedding method combined with self-attention mechanism is proposed to extract channel features from multiple domains simultaneously. This ``one model to fit them all'' solution employs available wireless channel data as the only data set for self-supervised pre-training. With the permission of users, network operators or other organizations can make use of some available user specific data to fine-tune this pre-trained realistic channel model for applications on channel-related downstream tasks. Moreover, even without fine-tuning, we show that the pre-trained realistic channel model itself is a great tool with its understanding of wireless channel.
\end{abstract}


\section{Introduction}
Channel modeling is essential for the wireless system design. Currently, stochastic channel models are widely used \cite{kermoal2002stochastic}. It is relatively easy for a stochastic model to generate completely consistent wireless channels for efficient calibrations among different algorithms. However, stochastic models are too generic to provide enough details for a certain scenario, resulting in some misleading conclusions. In spite of its behavioral stochastic property, a true wireless channel results from some deterministic propagations. Given an exact 3D map of a radio propagation environment, one can combine reflections, refractions, diffractions, shadowing and wave-guiding phenomena of the radio wave, perform a ray-tracing method, and finally predict a deterministic channel model \cite{hur2016proposal}. This environment dependency makes a deterministic channel model more precise than a stochastic one for a predication in a certain scenario. Admittedly, it¡¯s arduous to restore every piece of detailed information of time-varying environment and of hundreds of radio paths.

Instead of stochastic channel models purely built on stochastic properties and deterministic channel models built on geometrical environment of wireless channels, we propose a realistic channel model to reach a similar accuracy as deterministic channel models and uniformity as stochastic channel models. This model has three advantages. Firstly, it is a one-step model learned from wireless channel and used for wireless channel tasks. The environment is learned as latent information. Secondly, it is a unified ``one model to fit them all'' solution for various wireless channel tasks. Thirdly, it protects personal data without labeling data or privacy data as channel data is the only input. Bringing intelligence in physical layer combined with the smart network optimization \cite{chuai2019collaborative} will surely become a major source of gain in beyond 5G era. The proposed realistic channel model can help both. For example, a good comprehension and predication of current channel status can help radio devices to schedule, form beams, and to control power in a more efficient way.

For these advantages, we use a self-supervised pre-training approach to model wireless channel characteristics. Self-supervised learning is an autonomous supervised learning system that can generate an infinite number of labeled samples automatically. Most precedent works on machine learning-based channel model is specific: train a neural network based model with a specific wireless channel data and apply it for some specific channel tasks, such as channel feedback \cite{wen2018deep}, localization \cite{decurninge2018csi}, and channel prediction \cite{arnold2019enabling} and so on. However, it is inefficient and time-consuming to have one model for one channel-related task and to retrain it totally from scratch if applied to another. To solve it, we introduce a pre-training trick that has been widely used in computer vision (CV) and natural language processing (NLP). This trick can allow our model to learn some inherent structural properties from input data and then to establish the stage for the ensuing task-specific fine-tuning phase. As far as we know, it is the first time to realize self-supervised pre-training of wireless channel and yield a realistic channel model from it. With this pre-trained realistic channel model, network operators could adjust the models with modest re-training effort and volume of user-specific privacy data, and then apply them to downstream wireless channel related tasks such as network optimization.

A pre-trained realistic channel model can be used as a starting point for a wide variety of channel related tasks, and has two major capabilities. First is ``comprehension". It can score any input channel based on how likely it occurs in the current environment. It can even output an equivalent version of the input channel according to its own understanding of it. Second is ``prediction". It allows us to generate new channels by extending or replacing the input channel based on the computed probabilities. For example, predicting or interpolating (extrapolation) channel is to extend current channel while prediction on an existed channel based on its context can be used to replace this channel. The predication can be cross domain: interpolation or extrapolation on the frequency domain can be used for predication on the time domain and vice versa. The main contributions of this work are as follows:

\begin{itemize}
\item A proposal and implementation of a realistic channel model, which is pre-trained with channel data in a self-supervised manner. It is strong enough to freeze channel changing rules within itself, meanwhile, it is simple enough to support channel related tasks as stem supporting head.
\item The first demonstration of combining task related multi-domain features and self-attention mechanism of Transformer architecture. This will simplify the model design when dealing with data having multiple domains.
\end{itemize}

\section{Preliminary}
\label{section:preliminary}
In this section, we introduce two techniques to process the channel data and explain why using them is beneficial and necessary.
\subsection{Self-attention mechanism}
\label{subsection:self-attention}
Recurrent neural network (RNN) is a good start to deal with time series tasks by sequentially processing each input. However, a wireless channel is sampled in the multiple domains rather than a single time domain, it is no longer appropriate to sequentially process each input with RNN. To establish a suitable model for a realistic channel, the internal relations of input channel should be taken care. Self-attention mechanism is capable of attending to the internal elements of an input.

\begin{figure*}[!htp]
\setlength{\abovecaptionskip}{0pt}
\setlength{\belowcaptionskip}{0pt}
\centerline{\includegraphics[width=5in]{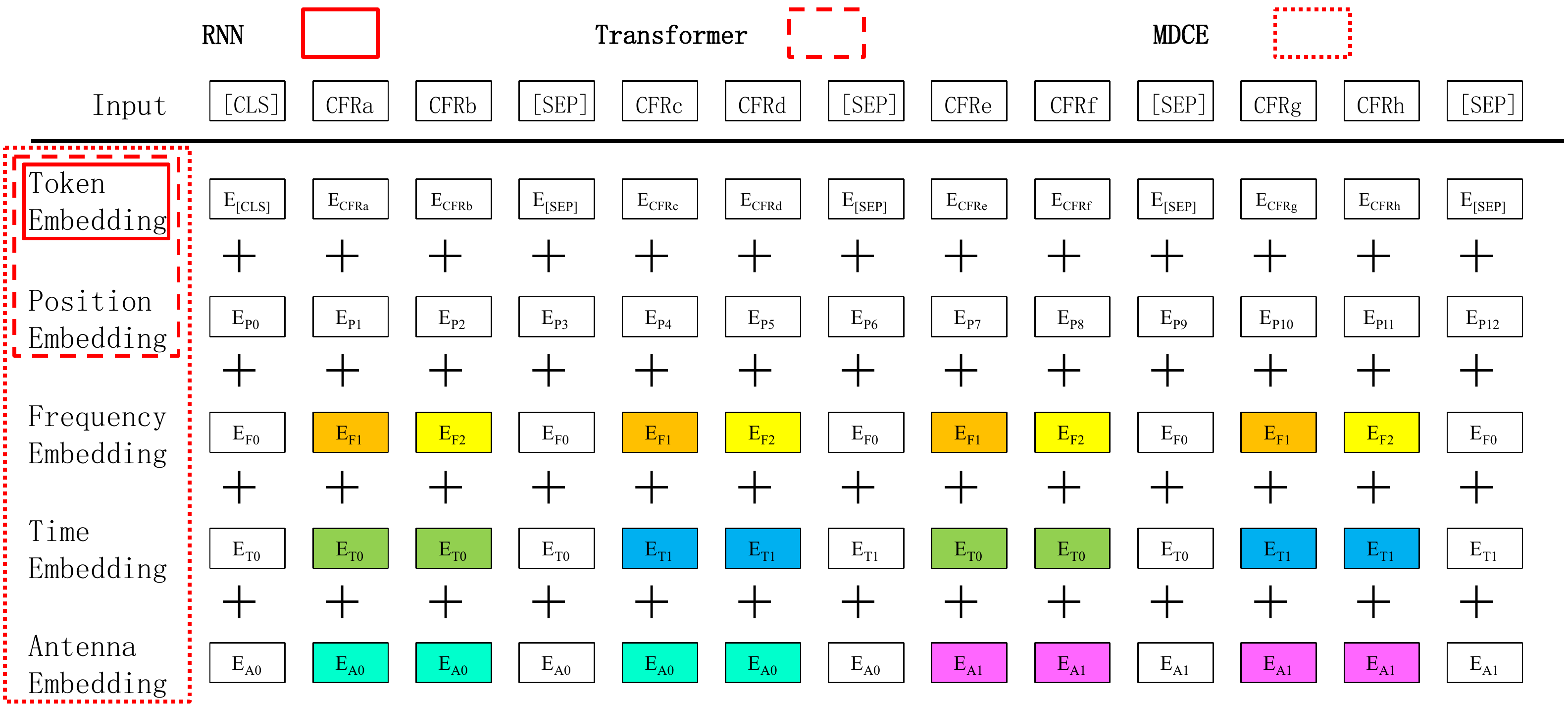}}
\caption{Schematic diagram of multi-domain channel embedding (MDCE) and its difference from RNN and Transformer based embeddings.}
\label{fig:mdce}
\vspace{-0.5cm}
\end{figure*}

Self-attention mechanism is proposed along with Transformer architecture \cite{vaswani2017attention} that is more computationally attractive than RNN attributed to the parallel processing properties. Moreover, RNN has difficulties to remember the long-term information, which makes sense for language model, because the closer words have more relevance to predict the next. However, it makes less sense for channel data, as the multi-domain property and underlying environment information complicates the correlations of channel input. If RNN was for channel input, we would assume that a closer input data be more important. In contrast, Transformer based model does not assume that. Instead, it learns out the relevance between channel input and output from an equal start by employing self-attention mechanisms. For a complicated problem, we tend to believe that any prior assumption is arrogant and may be misleading in later trainings. Therefore, it is wiser to start equally and leave it to neural network than start with priority.

\subsection{Channel Embedding}
\label{subsection:embedding}
In Orthogonal Frequency Division Multiplexing (OFDM) system, where complex-valued channel frequency response (CFR) is explicitly estimated but implicitly contains many others such as multipath propagation, Doppler Effect and interferences. To dig out these high dimensional characteristics, a high dimensional representation should be adopted. In the previous work \cite{huangfu2019predicting}, we have managed to extract some sequential features of a channel by using channel embedding technique, and fulfilled channel prediction task. An embedding of wireless channel is a mapping of a channel coefficient to a high dimensional vector with learnable numbers. In another word, we treat the channel input as categorical data for classification task, assuming that there are V possible values of channels, i.e., a size-V vocabulary, with which we can turn each channel into a W-length embedding for training. We pre-process the channel data in this way for three reasons.

First, it is confusing to separate a complex signal into two parts for neural network computations. We are wondering how much information about the physical channel would be lost due to this separation. Moreover, simple inputs may become a bottleneck for digging out underlying rules as the superficial rules are too easy to get. Secondly, lower power of the input channel impacts weights of the model less, it means these weak channels will almost be treated as if they did not exist, which is obviously not right. Normally, batch normalization can be used to overcome this imbalance by scaling the input. However, scaling powers of weak channels to a higher value is confusing, features of which cannot be learned. As for embedding representation, channel with different powers can have equal weights on learning models and results. Thirdly, it¡¯s possible to classify the channels into V possible values in terms of its statistical behavior and still keep the channel accuracy in a reasonable range. Nowadays, neural network models can handle a vocabulary size V as large as half million.

\section{Realistic channel model}
\label{section:realistic}
\subsection{Multi-Domain Channel Embedding}
\label{subsection:multi-domain}

To support various channel tasks, we should extract all features of a wireless channel. To deal with the wireless channel in the time, frequency, and spatial domains, we propose a method called as multi-domain channel embedding (MDCE) that simultaneously extracts the features in the multiple domains. As shown in Fig. \ref{fig:mdce}, the input is estimated CFR represented by token embeddings as discussed in section \ref{subsection:embedding}. For Transformer-based models such as BERT \cite{devlin2018bert} or GPT \cite{radford2019language}, if the input is in parallel, position embeddings should be added to the token embeddings to inject the positional information. If the input is a time series, position embeddings could learn the time relations. However, considering a wireless link between a base station and a moving user, the channel varies not only over time, but also over frequency and space. If we mixed up these multi-domain data, the model would be confusing. In contrast, we can add these channel features in the multiple domains as the additional embeddings, and leave the model to learn them, i.e., MDCE learning process. Raw channel data has three basic domains (spatial, time and frequency), features of which can be obtained without manually labeling the data. Using MDCE for training has two benefits:

First, it would allow the model to learn channel correlations on these domains and yield a better understanding about the environment. For example, in a massive MIMO system, as some antennas are correlated in near-field propagation while others not, antenna embeddings can help on extracting these features of interests.

Second, it enables the model to predict channel on arbitrary domain, which means more channel related tasks can be achieved with this model. For example, we can predict channel on some subcarriers based on other subcarriers, or we can predict future channel based on the current channel.

Fig. \ref{fig:mdce} shows a schematic of MDCE with three domains. Limited by the figure size, for every domain the number of features is set to 2. Note that in a realistic training procedure the inputs could be a long sequence and number of features large. For frequency embeddings, E$_{F1}$ and E$_{F2}$ correspond to two subcarriers. For time embeddings, E$_{T0}$ and E$_{T1}$ could represent current sub-frame and next sub-frame with a certain time interval. For antenna embeddings, E$_{A0}$ and E$_{A1}$ indicate two antennas used for receiving or transmitting the signals. Note that the frequency embedding could be replaced by delay embedding if the channel is transformed into delay domain. Labeled data from other domains can also be added, such as user speed, cellular ID, user equipment ID and so on. However, introduction of non-channel labeled data during pre-training is out of the scope of this paper. There are two special tokens in the inputs, ¡®[CLS]¡¯ and ¡®[SEP]¡¯, the first token can be used to classify or represent the input, and the second one is a separation token for clear observation.

\subsection{Intra-domain self-attentions}
\label{subsection:intra-domain}
\begin{figure}[!tp]
\setlength{\abovecaptionskip}{0pt}
\setlength{\belowcaptionskip}{0pt}
\centerline{\includegraphics[width=3.3in]{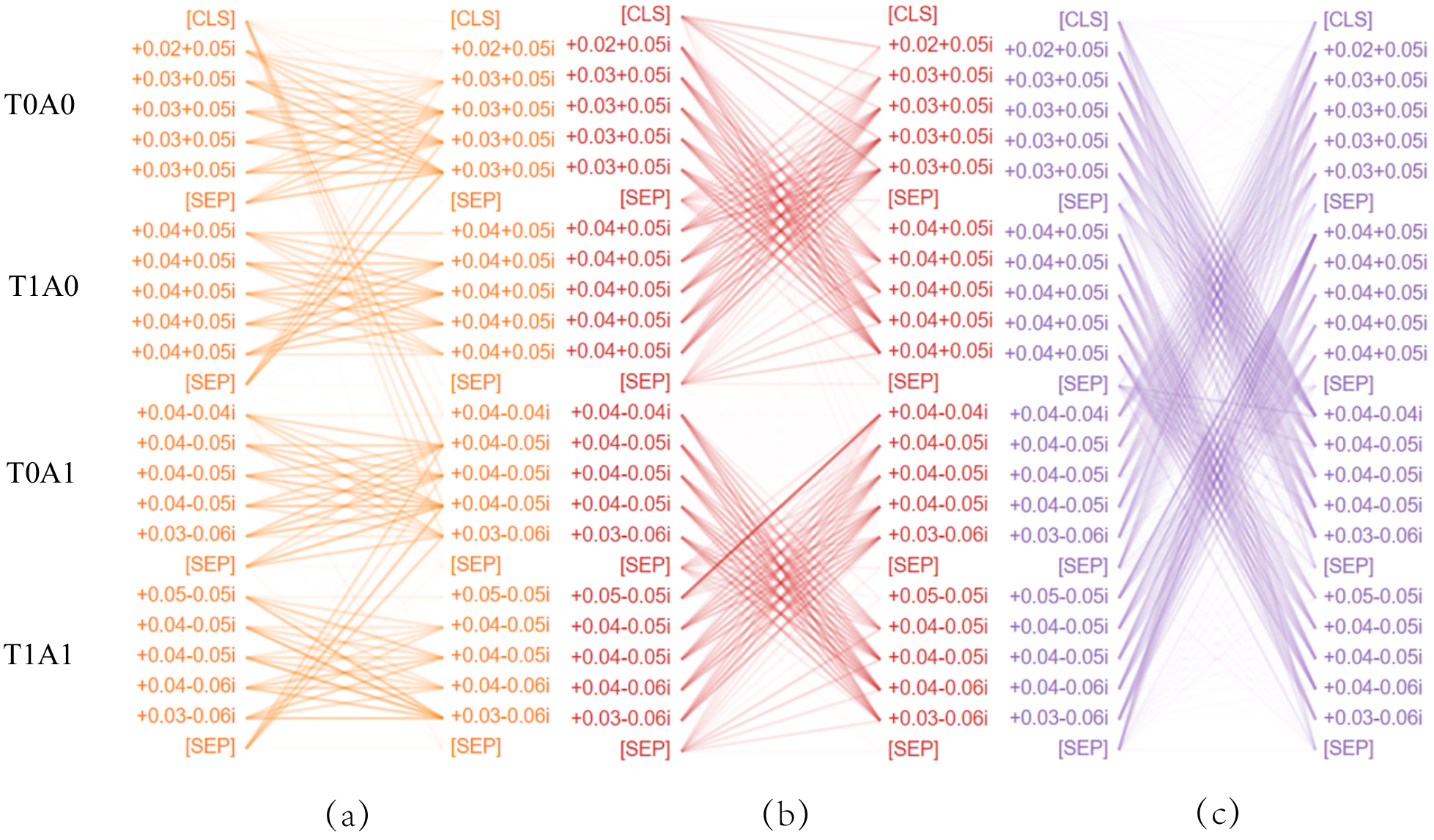}}
\caption{Self-attentions of channel input in multiple domains with a pre-trained channel model.}
\label{fig:self-attentions}
\vspace{-0.6cm}
\end{figure}

A channel input sequence is composed of channel subsequences in multiple domains, and each subsequence has a fixed combination of domains. We believe that MDCE can help self-attentions focus on intra-domain channel correlations. As a result of combining MDCE and self-attention mechanism for pre-training, we find that some self-attentions start to show this intra-domain behavior. During pre-training, multiple attention patterns are trained simultaneously. From the patterns shown in Fig. \ref{fig:self-attentions}, we can sense the mechanisms of self-attentions.

We herein provide an example model that is pre-trained with many channel samples on 200 subcarriers, 2 consecutive time frames (T0, T1), and 2 antennas(A0, A1). For a convenient visualization, we use a visualization tool \cite{vig2019multiscale} and keep only the first 5 subcarriers on the frequency domain. The tool visualizes attentions as some lines connecting the position being updated (left) with the position being attended to (right), and the thickness of the lines reflecting their attention scores. Pattern in Fig. \ref{fig:self-attentions}a shows that the attentions are mostly directed to surrounding subcarriers from the same time frame and the same antenna. This attention pattern occurs frequently and demonstrates that frequency correlations are useful on the prediction of the current channel. In Fig. \ref{fig:self-attentions}b, attention seems to be directed to the other time frame and the same antenna, while attention in Fig. \ref{fig:self-attentions}c is mainly on the other antenna and the same time frame. These attentions correspond to time correlations and spatial correlations respectively that are all learned by the model itself thanks to the MDCE method.

\subsection{Transferring}
\label{subsection:transferring}
\begin{figure}[!t]
\setlength{\abovecaptionskip}{0pt}
\setlength{\belowcaptionskip}{0pt}
\centerline{\includegraphics[width=3in]{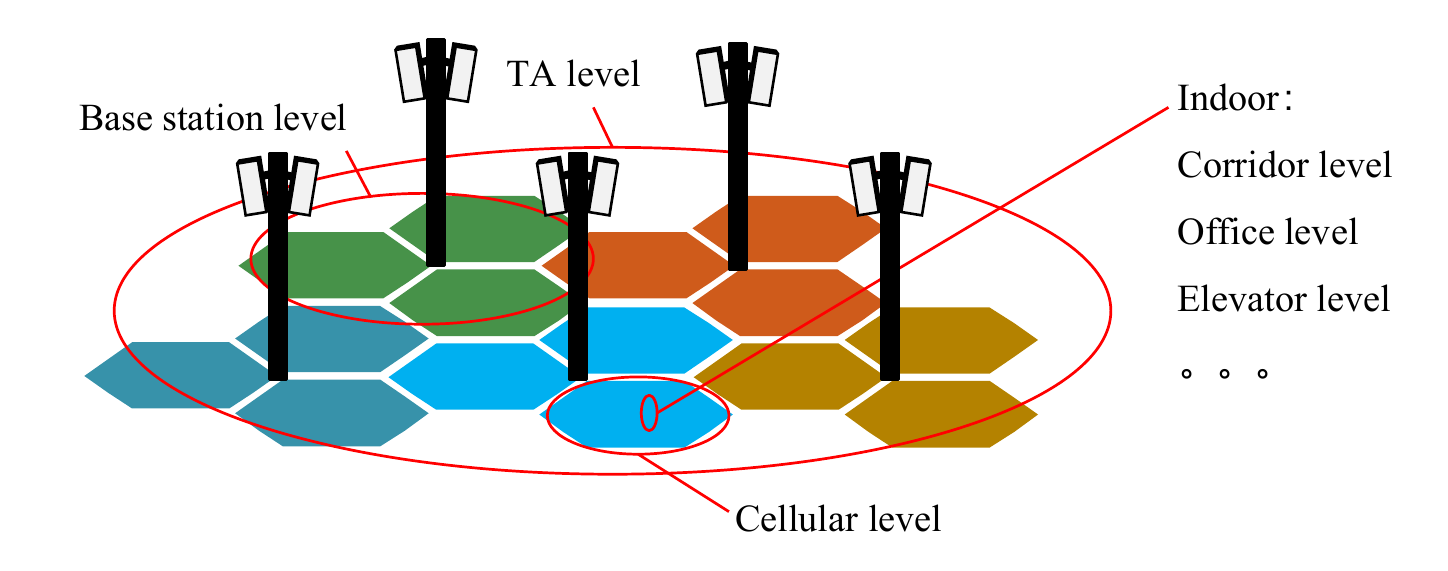}}
\caption{Schematic diagram of realistic channel models in different levels.}
\label{fig:level}
\vspace{-0.6cm}
\end{figure}

In some realistic scenarios, to help physical layer computations and network optimization, the pre-trained channel models can be trained and deployed at different geographic levels. As shown in Fig. \ref{fig:level}, the geographic levels could be tracking area (TA), base station and cellular for outdoor scenario; for indoor scenario, the levels could be office, corridor, elevator, and so on. Strictly speaking, there is no identical environment. However, it is believed that neighbor realistic channel models in the same geographic level tend to have similar features, e.g., different offices may have similar layouts. Therefore, a pre-trained office channel model could be transferred among wireless devices of offices of the kind and still ready for directly use, which reflects the uniformity of the proposed channel modeling method.

Normally, a pre-trained channel model is carefully trained with high-quality and large-volume data. Not all wireless equipment has the ability to pre-train an accurate robust channel model. Moreover, using a transferred pre-trained channel model rather than training from scratch can save remarkable computing and time resources. Transferring a pre-trained channel model from source device to target device can be followed by transfer learning with channel data from target device. However, at target device, the distribution of pilot signal strength caused by power control strategy or other settings may be different from source device so that the channel data to be learned has a different scale from the training data of pre-trained model. Therefore, a scale finding algorithm must be used to reduce the gap between pre-trained channel model and data from target device. If an optimal channel scale was found, the channel input should become reasonable from perspective of the model. In other words, the perplexity of channel input should be low. For channel input C with sequence length N, there is
$C=(C_1,C_2,...,C_N)$,
The likelihood of this channel sequence from perspective of the channel model is as follows,
$P(C)=P(C_1,C_2,...,C_N)=\prod_{i=1}^{N}P(C_i|context)$,
Where context could be the preceding channel data $(C_1,...,C_{i-1})$ for unidirectional model such as GPT, and for bi-directional model like BERT context would be the surrounding channel data $(C_1,...,C_{i-1},C_{i+1},...,C_N)$, and the perplexity of this channel sequence would be,
$PP(C)=P(C)^{-\frac{1}{N}}$,
With channel scale $S$, channel $C$ normalized by $S$ would be,
$CS=(C_1,C_2,...,C_N)/S$,
The algorithm we proposed is to find the optimal $S$ to minimize the perplexity of input channel,
$S=\argmin_{S}PP(CS)$,
which is as follows,
\begin{algorithm}[h]
	\caption{Channel Scale Finding For Transferring}
	\label{alg:finding}
	\begin{algorithmic}
        \State Get channel $C$ and channel model $M$
		\State Initialize normalizer $S$
		\For {each $i \in [1,IterNums]$}
		\State Compute $CS$ with $C$ and $S$
		\State \Call{PP}{$CS, M$}
        \State Record pair of $(S, PP)$
        \State Assign new $S$
		\EndFor
		\State Find $S$ with the smallest $PP$
		\Function{PP}{$C, M$}
		\State Compute the perplexity of $C$ with model $M$
		\EndFunction
	\end{algorithmic}
\end{algorithm}

After the optimal channel scale $S$ is found, the channel data from target device is normalized with $S$, and transfer learning could be performed on the transferred model. This operation will further update the parameters of the model to fit the distribution of the new channel data. Except the channel scale, source device should also transfer a mapping table that records the feature types corresponding to MDCE to target device so that the target device can correctly pre-process the channel data.

\subsection{Fine-tuning with channel related labels}
\label{subsection:fine-tuning}
After self-supervised pre-training of a realistic channel model, a small volume of channel related data with labels can be used to tune the model. To fine-tune on a specific task, we can use the final hidden states of special token ¡®[CLS]¡¯ as the aggregate representation of inputs. This representation can then be introduced into a newly added classification layer for supervised learning. By training with the loss between the labeled data and the predicted classification results, the weights from pre-trained model and the new layer are jointly updated to maximize the log-probability of the correct labels. Afterwards, a task-specific model can be achieved. Any CSI related task could be a downstream task, including (but not limited to) localization, beamforming, user pairing, network optimization and so on.

\section{Evaluation setup}
\label{section:setup}
This section describes the pre-training setup for realistic channel model.

\subsection{Datasets}
\label{subsection:datasets}
The channel data is collected in outdoor area of Shanghai Waitan with a walking speed. Downlink Cell Reference Signal (CRS) is measured for commercial LTE system with 2 by 2 antenna. The system bandwidth is 20MHz, i.e., 100 resource blocks (RB). In one sub-frame, 200 subcarriers multiplies 4 symbols are used for pilots allocation for each port. The carrier frequency is at 1.9GHz and CFR is estimated from pilot symbols. The least-square estimated channel is truncated at time domain while distinctive taps are kept for de-noising processing. The channel data used herein is from a wireless link between our measuring equipment and a base station and lasts for 280 seconds. The raw file is about 6.6GByte. Total number of CFRs is around 0.9 billion. Channel data from antenna 0 of user is used for train set, while data from antenna 1 is for test set. To transform the complex-valued channel into categories, two significant digits are introduced to keep the precision in a reasonable range. Afterwards, by counting the number of occurrences for each channel category, we can select the top V frequently occurred channel and put them into a channel vocabulary with size V. V is about 18000 in this case.

\subsection{Architecture}
\label{subsection:architecture}
The model size of BERT$_{BASE}$ is adopted, there are 12 layers, and 12 attention heads for each layer, hidden size is 768, the total number of parameters is 110M. The customized MDCE is summated up and the model was trained from scratch for channel modeling. Each batch contains 12 sequences, each sequence contains 800 channel data (200 subcarriers $\times$ 2 time frames $\times$ 2 antennas) , and 5 special tokens are used to divide the sequence into 4 subsequences. Three epochs are utilized for pre-training. Learning rate starts at $5e^{-5}$ with a warm-up and a linear decay procedure. For pre-training, two prediction tasks similar as BERT are used to train the channel model. One is to predict some masked channels while the other one is to predict if the channel are temporal consecutive called as next time frame prediction task.

\section{Results and discussions}
\label{section:results}
The pre-trained model itself is a great tool even without fine-tuning. Let us see what it can do with its understanding of realistic channel.

\subsection{Mitigating pilot contamination}
\label{subsection:contamination}
Pilot contamination is a serious issue in a massive MIMO system due to pilot reusage across neighboring cells. It has been investigated that with the presence of pilot contamination, SINR saturates due to corruption of channel estimate by interfering user. It will impact an achievable rate and capacity \cite{elijah2015comprehensive, marzetta2010noncooperative}. To mitigate pilot contamination, we can use a two-step solution enabled by the pre-trained realistic channel model.

\begin{figure}[!t]
\setlength{\abovecaptionskip}{0pt}
\setlength{\belowcaptionskip}{0pt}
\centerline{\includegraphics[width=3in]{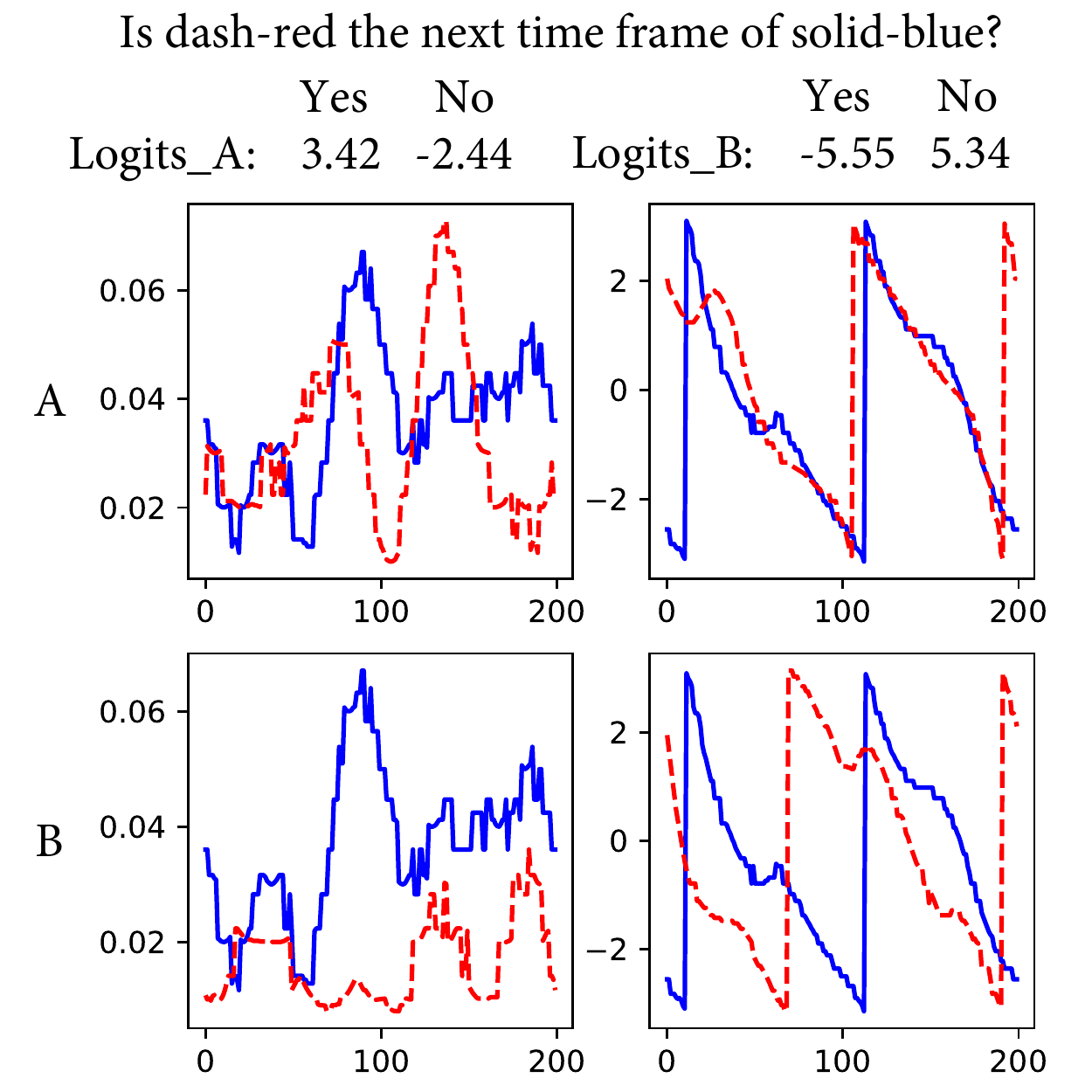}}
\caption{An example of predicting on next time frame task, groundtruth of case A and B is Yes and No respectively, solid-blue is the first time frame and dash-red is the second time frame, left figures are powers of channel and right figures are angles of channel.}
\label{fig:contamination}
\vspace{-0.5cm}
\end{figure}

The first step is to detect pilot contamination. In the train set, as the pilot contamination is far rarer than normal channel estimation, the model should have a good guess on whether the second time frame is the following frame. In our experiment, the accuracy of this next time frame prediction task is 98.6\%, high enough for practical use. For every two time frames input, if the output of model shows that the two time frames are non-consecutive, it is probably the case that pilot contamination happened. In Fig. \ref{fig:contamination}, we select two consecutive time frames for case A; while a random time frame is selected as the second time frame for case B. From the power (left) and phase (right) perspective, we can only tell there are some differences of first time frame (solid-blue line) and second time frame (dash-red line) in both cases. However, the pre-trained channel model can give not only the correct results but also a soft information, i.e., logit, related to how surely it is about this decision. Larger positive logit corresponds to higher probability. During training, these logits will be passed to softmax function for the computation of loss.

If the model is quite sure about the abnormality of channel from the next time frame, using this channel for following decoding and interpolating would be misleading. Then the second step is to generate a predicted channel to replace this one. We can simply mask the next time frame channel. Then the model will give its best guess for every masked token, as the model has seen and memorized a lot about the temporal correlations. The predicted result should be more reasonable than the contaminated one.

\subsection{Channel compressing}
\label{subsection:compressing}
To increase spatial multiplex and eliminate the interference among users, it is necessary for the transmitter to get an accurate and timely CSI. In frequency division duplexing (FDD) system, the receiver has to feedback CSI to transmitter, an overhead for massive MIMO due to the increasing number of antennas. Conventional channel compressing method presumes the channel as a sparse signal and compresses the sparse information via random projections. New attempt with neural networks such as CsiNet \cite{wen2018deep} used a CNN-based encoder to translate the channel into a codeword through its intermediate extracted features.

The purpose of channel compressing is to extract a short and distinctive representation from the input channel. As a matter of self-attentions, the first token ¡®[CLS]¡¯ attends to all the information of the input sequence when passed through the pre-trained realistic channel model. Therefore, its output states of the model¡¯s final hidden layer can be used as a fixed-dimension pooling representation of inputs. This representation is an H-length vector, where H is the hidden size of the model. For example, supposing that the input sequence is channel in several domains, the length of input to be compressed would be Ns*Nt*Nr*Nc*Nb, where Ns is number of subcarriers, Nt is number of antennas of transmitters, Nr is number of antennas of receivers, Nc is 2 as channel is in complex domain and Nb is the batch size to combine the representations from multiple inputs, we have a compression ratio $\gamma$ as the following equation,
$\gamma = N_s \times N_t \times N_r \times N_c \times N_b/H$.
In a specific scenario, the hidden size H should be fully investigated to find an optimal value, not only small enough to decrease the model size and increase the compression ratio but also be large enough to distinguish any input channel. It is very likely that the optimal H is related to the sparsity of current multipath channels \cite{bajwa2010compressed}. At the transmitter side, another model can be trained to map the representation into a corresponding beamforming or user pairing strategy. With the current settings in this work, the compression ratio is 25.

\subsection{Channel fingerprinting}
\label{subsection:fingerprinting}
Channel fingerprinting is to capture the geometry related channel features of an area \cite{ye2017neural}, the channel fingerprints have a property that the fingerprints corresponding to locations close in geometrical space should also be close in Euclidean space and vice versa. To achieve this, we can also use the aforementioned final hidden states of first token as fingerprints to distinguish channel in a fine granularity of milliseconds. The environment information would be learned as latent information within the high dimensional vector of these channel fingerprints.

To smooth the output, we take the average of 8 temporal consecutive sequences as a fingerprint, which means that each fingerprint represent the channel in 2 milliseconds, within this short duration, as coherent time. These fingerprints are ready for further downstream tasks such as localization if geometrical data is available. Meanwhile, a deeper look into the channel fingerprinting is channel charting \cite{studer2018channel} visualizing the fingerprints in a low-dimension space. As a channel fingerprint is an H-length vector, to visualize the fingerprints, we should generate a low-dimension representation of the fingerprints with a dimensionality reduction tool. Herein, we use t-Distributed Stochastic Neighbor Embedding (t-SNE) \cite{maaten2008visualizing} to reduce the dimension of fingerprints from 768 to 2.

\begin{figure}[!t]
\centering
\subfloat[]{\includegraphics[width=1.65in]{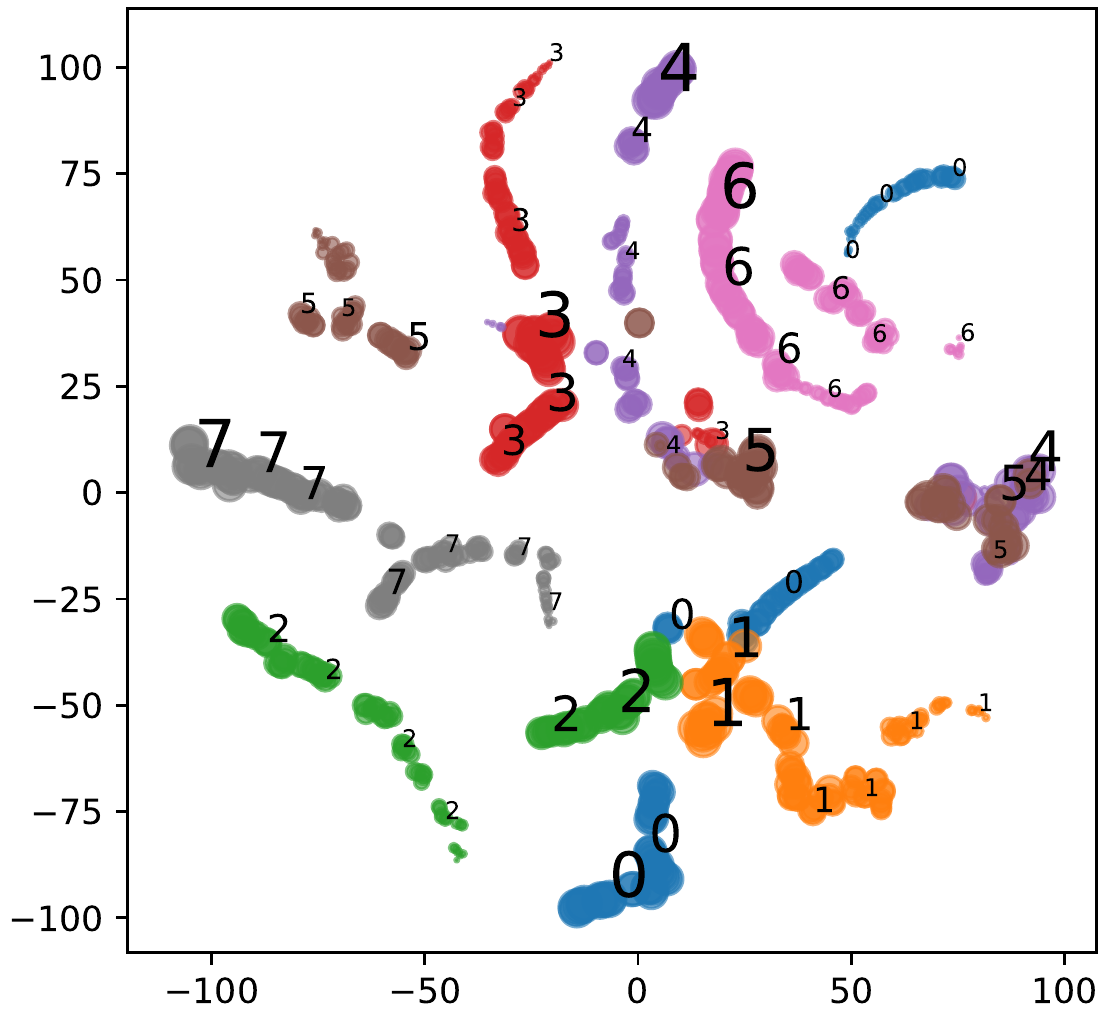}
\label{fig:2500_5}}
\subfloat[]{\includegraphics[width=1.65in]{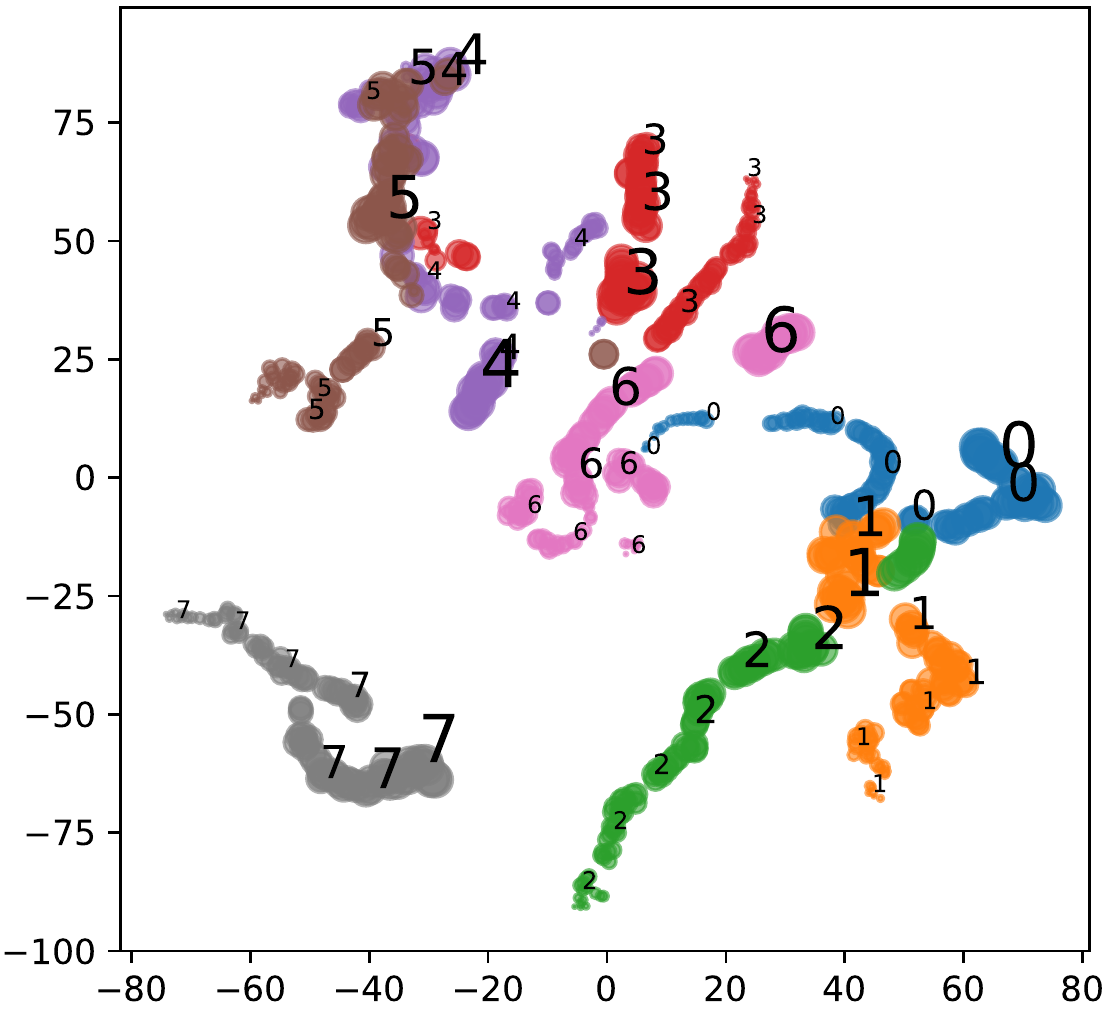}
\label{fig:2500_10}}
\caption{Channel charting of low-dimension representations of channel fingerprints using t-SNE with perplexity equals to (a) 5 (b) 10.}
\label{fig:charting}
\vspace{-0.5cm}
\end{figure}

As shown in Fig. \ref{fig:charting}, we pick 100 continuous fingerprints, i.e., 20 milliseconds every 5 seconds for 40 seconds, giving us 8 sets of fingerprints labeled with 0 to 7. Following the time flows during each 20 milliseconds, the size of labels and markers is increasing. In this chart, we can find that when user is moving, the fingerprints are also slowly changing and form several routes. Perplexity in t-SNE is a parameter corresponding to number of nearest neighbors k that is employed in many manifold learners. With 5 and 10 perplexity, we can view different-angle projections of manifold of channel fingerprints. For example, we can observe a broken route from label 2 in Fig. \ref{fig:charting}a while it becomes continuous in Fig. \ref{fig:charting}b. Meanwhile, the 8 routes are well separated, which is consistent with the mutation of channel features with 5 seconds interval. It shows that the geometrical information is well learned, even more, can be preserved in these low-dimension representations.

\section{Conclusions}
\label{section:conclusions}
This paper proposes to pre-train a realistic channel model in a self-supervised manner, which can be used as a uniform model for most channel related tasks. The training strategy introduces self-attention mechanism and multi-domain channel embedding to extract channel features. For practical usage, we also show an algorithm to transfer the pre-trained model from one device to another. From the results of using pre-trained model on several channel related tasks, we can find that pre-trained model has a good understanding of the current environment, and further downstream tasks can be fine-tuned with labeled data.

In future works, more downstream tasks should be designed to verify the performance of this model. For practical use, tradeoff between the model size and its performance should be investigated.

\bibliographystyle{IEEEtran}
\bibliography{PT1}
\end{document}